# ARE ALL CREDIT DEFAULT SWAP DATABASES EQUAL?

**Sergio Mayordomo**
*School of Economics and Business Administration. Universidad de Navarra, Spain.*
*Email:smayordomo@unav.es*

**Juan Ignacio Peña**
*Department of Business Administration, Universidad Carlos III de Madrid, Spain*
*Email: ypenya@eco.uc3m.es.*

**Eduardo S. Schwartz**
*Anderson Graduate School of Management, UCLA, CA, USA*
*Email:eduardo.schwartz@anderson.ucla.edu.*

## Abstract

*We compare the five major sources of corporate Credit Default Swap prices: GFI, Fenics, Reuters, CMA, and Markit, using the most liquid single name 5-year CDS in the iTraxx and CDX indexes from 2004 to 2010. Deviations from the common trend among prices in the different databases are not random but are explained by idiosyncratic factors, financing costs, global risk, and other trading factors. The CMA quotes lead the price discovery process. Moreover, we find that there is not a full agreement among databases in the results of the price discovery analysis between stock and CDS returns.*

**Keywords**: *Credit Default Swap prices, Databases, Liquidity*

**JEL classifications**: *F33, G12, H63*

This paper was partially drafted during the visit of Sergio Mayordomo to the Anderson School at UCLA. We acknowledge financial support from MCI grant ECO2009-12551. We thank Rodolfo Campos, Teresa Corzo, Lars Norden, Yves Nosbusch, J. Pedro Nunes, Jose Penalva, Maria Rodriguez-Moreno, Christina Wang, and other participants in the IX INFINITI Conference, 2011 IFABS Conference, XIX Foro de Finanzas, University of Valencia seminar, and 2012 EFMA Conference (Barcelona) for useful comments. An anonymous referee provided many astute suggestions that considerably improved the paper. Corresponding author: Sergio Mayordomo.

# 1. Introduction

Over the last decade, the Credit Default Swap (CDS) market has grown rapidly.[1] Given the growth and the size of this market, quoted and transaction prices of CDS contracts are widely thought to be a gauge of financial markets' overall situation, as suggested by the GM/Ford credit episode in 2005, the US subprime fiasco in 2007-2009 or the Europe's debt crisis in 2010. Academic and policymakers alike have voiced concerns with respect to the CDS market's role in the above mentioned episodes and its possible influence in other financial markets, credit-oriented or otherwise. However, to properly address current concerns, careful empirical research is needed and therefore dependable CDS price data is a key requirement. The CDS market is an Over the Counter (OTC) market almost entirely populated by institutional investors and therefore, in contrast with an organized exchange like the NYSE, there is no reliable information on prices. The information on prices must be gathered from market participants on the basis of their voluntary participation on periodic surveys, with all the potential shortcomings such a situation may bring about. For instance, Leland (2009) reports that Bloomberg's CDS data is frequently revised weeks after and often disagrees substantially with other data sources such as Datastream. Given that price data deserve special attention, as the validity and power of the empirical results must be based on a dependable data source, in this study we investigate the differences in the main data sources employed by researchers and policymakers in this area. Specifically, we compare the five data sources for CDS prices commonly used in almost all the extant research: GFI, Fenics, Reuters EOD, Credit

---

1 The global notional value of CDSs outstanding at the end of 2004, 2005 and 2006 was $8.42, $17.1 and $34.4 trillion, respectively. The CDS market exploded over the past decade to more than $45 trillion in mid-2007 and more than $62 trillion in the second half of the same year, according to the ISDA. The size of the (notional) CDS market in mid-2007 is roughly twice the size of the U.S. stock market (which is valued at about $22 trillion) and far exceeds the $7.1 trillion mortgage market and $4.4 trillion U.S. treasuries market. However, the notional amount outstanding decreased significantly during 2008 to $54.6 trillion in mid-2008 and $38.6 trillion at the end of 2008. This declining trend followed in 2009 (31.2 in mid-2009 and 30.4 at the end of 2009), but seems to have stabilized in 2010 and 2011.

Market Analysis (CMA) DataVision (CMA hereafter), and Markit. We study the consistency of these five CDS data sources in the cross section and time series dimensions using the most liquid single name 5-year CDS of the components of the leading market indexes, iTraxx (European firms) and CDX (US firms). First we look at their basic statistical properties. Then we address two specific issues: (i) the factors explaining the divergences from the common trend among different CDS quoted spreads, and (ii) the relative informational advantage of the prices coming from different CDS databases. Finally, we study the consistency among databases in the results of a price discovery (causality) analysis between stock and CDS returns.

Two price time series for the same single name CDS reported by different data sources should, in principle, be very close in the sense that both share a common trend, the underlying true value of the asset. Even if there are deviations from the common trend between the price series reported by the different datasets, one should expect that these deviations are random errors and therefore unrelated both to idiosyncratic factors such as firm size, the volatility of the firm equity prices, or the disagreement/agreement of analysts' forecasts on the firms earnings, and to systematic factors such as the financing costs of the participants in the CDS market, global risk, or trading activity. If all the data sources are consistent among them, the use of a given data source should not affect research results and their financial and policy implications. But, if there are significant deviations among them, the research implications may be sensitive to the specific data base employed. Any inconsistency in prices from private providers would also have implications for market transparency which would affect all financial agents such as investors, risk managers, and regulators.

We find that there are systematic departures from the common CDS spreads' trend across databases. Our analysis suggests that, although the different CDS quotes move broadly together, there are very noticeable divergences for some entities in some days. Also, the discrepancies among databases appear to be more marked in specific time periods, probably reflecting market turbulences. No single database, however, provides quotes that are consistently above or below the quotes from other databases. We also find evidence suggesting that on average the days without trade price information have higher quote dispersion than the days with trade price information.

Most importantly, deviations (in absolute value) from the common trend among the different CDS quoted spreads are not purely random, but are related to idiosyncratic factors such as the disagreement of analysts' forecasts of the company's earnings per share (EPS) or the firm size, and also to specific CDS liquidity, global risk, global financing costs, and trading factors. We also find that the different data sources do not reflect credit risk information equally efficiently. Our results suggest that, for the sample period considered, CMA quoted CDS spreads led the credit risk price discovery process with respect to the quotes provided by other databases.

The discrepancies among databases might have a material impact on the results of price discovery tests between stock and CDS returns given that we do not find a full consistency among databases in the results of the stock-CDS price discovery (causality) analysis.

Our results have a number of important implications for empirical research using CDS prices. First, for single-name CDS with low trade frequency, our results cast doubts on the reliability of the existing quoted price information. Second, the larger the disagreement in the analysts' forecast on the firms EPS and the smaller the firm, the larger are the deviations among databases. Third the higher financing costs of financial institutions and, the higher the global risk (measured by implied volatility indexes), the larger are the deviations from the common trend in prices across the different databases. Fourth, there is a CDS database that leads the price discovery process when they are compared to each other. Fifth, the price discovery analysis between stock and CDS returns reveals that there is not a full consistency among databases in terms of the causality relations found between stocks and CDS. Extensive robustness tests would be needed before a causality relationship between stock and CDS returns could be established.

This paper is structured as follows. Section 2 presents a literature review. Section 3 describes the data employed in the analysis. Section 4 motivates the research hypotheses and introduces the methodology. Section 5 shows the empirical results. Section 6 addresses the extent to which the differences among databases impact the price discovery tests between stock and CDS returns. Section 7 confirms the robustness of the results and presents some extensions. Section 8 concludes.

## 2. Literature Review

The importance of comparing alternative financial databases is stressed in the classical papers by Rosenberg and Houglet (1974) and Bennin (1980) on the differences between CRSP and COMPUSTAT stock price data. However, in more recent times there are very few papers comparing different databases. Schoar (2002) and Villalonga (2004) compare COMPUSTAT with the Longitudinal Research Database and the Business Information Tracking Series from the U.S Bureau of the Census, respectively, and show that different data sources have large impact on the answers to research questions. Despite the widespread use of CDS databases and the high relevance of their accuracy, to the best of our knowledge there exists no study that examines or compares data as well as databases. Our paper is a first attempt to fill this gap in the literature.

The first papers that compare, at least to some extent, different CDS data sources are Nashikkar and Subrahmanyam (2007) and Nashikkar, Subrahmanyam, and Mahanti (2009). However, this comparison is not the main focus of their paper and these authors do not present a detailed analysis. They simply conduct a test to ensure consistency between the CMA and GFI CDS spreads series over a short period when there was an overlap between the two series. They develop this test just to match GFI and CMA series and create a longer dataset given that they have the two data sources for different periods. Moreover, they do not report any results of the tests and simply state that they find consistency.

Mayordomo, Peña and Romo (2011) employ four different data sources (GFI, CMA, Reuters EOD and JP Morgan) to study the existence of arbitrage opportunities in credit derivatives markets focusing their attention to the single names CDSs and asset swaps. Although they find similar results employing any of the four previous data sources at the aggregate level, some differences appear at the individual reference entity level. They report their base results using GFI data but when they use the other data sources they do not find exactly the same number of arbitrage opportunities. For some individual firms they find arbitrage opportunities using GFI but they do not find them using some of the other data sources. In some other cases they find the opposite.

The Mayordomo et al.'s (2011) study above suggests that the differences in CDS prices from different databases can have a material influence on research results and therefore a careful analysis of the publicly accessible databases is called for. In fact, the problem could be potentially even more serious when researchers work with "unique" databases coming from a single dealer's quotes (contributor) and without crosschecking. In this study use a broad array of CDS data sources where, for most of them, prices are put together based on information provided by several market traders and dealers. Using aggregate prices we focus on the market factors or characteristics that could affect the consistency among quoted prices. Thus, instead of using individual dealer's prices we use aggregated (composite/consensus) prices which allow us to have a more comprehensive perspective on the market.

The only previous paper that employs different CDS prices (trades and quotes) is Arora, Gandhi and Longstaff (2012). They examine how counterparty credit risk affects the pricing of CDS contracts using a proprietary data set. Specifically, their data set spans from March 2008 to January 2009 and includes contemporaneous CDS transaction prices and quotations provided by 14 large CDS dealers for selling protection on the same set of underlying reference firms. The authors find differences across dealers in how counterparty credit risk is priced. That is, counterparty credit risk is not priced symmetrically across dealers and they consider that these asymmetries could be due to differences in the microstructure and legal framework of the CDS market. They argue that dealers may behave strategically in terms of their offers to sell credit protection.

We use aggregate data, which are formed after grouping the information of the market traders and dealers instead of individual dealer prices, to study the potential divergence among the composite CDS spreads. By concentrating on the aggregate prices we focus on the market factors or characteristics that could affect the consistency among quoted prices but we do not try to explain the effect of potential differences among the individual dealers. As Arora, et al. (2012) sustain, the decentralized nature of the CDS markets makes the transaction prices somewhat difficult to observe. This is why most empirical research analyses based on the CDS markets use price quotes instead of transaction prices.

Longstaff, Mithal, and Neis (2005) argue that the composite prices include quotations from a variety of credit derivatives dealers and therefore, these quotations should be representative of the entire credit derivatives market. We complement our analysis using also GFI transaction prices and Fenics prices (elaborated by GFI) which are based on a combination of transaction and judgmental prices, the latter computed using the Hull and White methodology and therefore not dependent on contributors.

## 3. Data

The five publicly available data sources that we employ in this paper are GFI, Fenics, Reuters EOD, CMA, and Markit.

- GFI, which provides traded CDS spreads, is a major inter-dealer broker (IDB) specializing in the trading of credit derivatives. GFI data contain single name CDS transaction prices for 1, 2, 3, 4, and 5 years maturities. They are not consensus or indicative prices. Thus, these prices are an accurate indication of where the CDS markets traded and closed for a given day. GFI data have been used by Hull, Predescu, and White (2004), Predescu (2006), Saita (2006), Nashikkar and Subrahmanyam (2007), Nashikkar, Subrahmanyam, and Mahanti (2009) among others.

- Fenics (elaborated by GFI) data are a mixture of traded, quoted and estimated CDS spreads. Fenics' data are credit curves for the whole term structure of maturities, generated hourly (all trading days) for more than 1900 reference entities. Data points in a given name's credit curve can be actual trades or mid prices calculated from the bid/offer quotes. If there are no market references, the Fenics CDS spread is computed using the Hull and White methodology to ensure that a credit curve always exists for each reference entity.[2] Fenics data have been used in Mayordomo et al. (2011) among others[3].

- Reuters EOD provides CDS composite prices. Reuters takes CDS quotes each day from over 30 contributors around the world and offers end of day data for single names CDSs. Before computing a daily composite spread, it applies a rigorous screening procedure to eliminate outliers or doubtful data. Mayordomo et al. (2011), among others, employ CDSs data from Reuters.

---

[2] Although Fenics is computed using the approximations mentioned above, it is a reasonably accurate data source. For instance, the median of the absolute difference in basis points between five years CDS premiums as defined by Fenics and the actual quotes or transaction prices registered in other databases for the period between April 2001 and May 2002, is equal to 1.16, 2.01 and 3.82 bps for AAA/AA, A and BBB ratings for a total of 2,659, 9,585 and 8,170 companies respectively.

[3] GFI is a broker which also reports the Fenics prices. The data reported by GFI are transactions prices or bid/quotes in which capital is actually committed. This data is only available when there is a trade. When there is not, GFI constructs the Fenics curve which is available daily with no gaps. To compute the Fenics curve, GFI uses its own information on transactions or quotes. If for a given day neither prices nor quotes are available, Fenics data is computed by means of Hull and White's methodology.

- Credit Market Analysis (CMA) DataVision does not categorize its CDS prices along the "composite" or "consensus" lines but in order to bring more transparency to CDS information, CMA uses a strict aggregation methodology, instead of "composite" or "consensus" methods, depending on the intraday market activity. The data aggregation is not equally weighted but the different weights are based on the respective age and length of the original sample employed (the last contribution is more influential than the older ones). CMA collects its CDS data from a robust consortium which consists of around 40 members from the buy-side community (hedge funds, asset managers, and major investment banks) who are active participants in the CDS market.[4] CMA reports bid, ask and mid prices. Among the papers that employ CMA data are Nashikkar and Subrahmanyam (2007) and Nashikkar, Subrahmanyam, and Mahanti (2009).

- Markit provides composite prices. The Markit Group collects more than a million CDS quotes contributed by more than 30 major market participants on a daily basis. The quotes are subject to filtering that removes outliers and stale observations. Markit then computes a daily composite spread only if it has two or more contributors. Once Markit starts pricing a CDS contract, data will be available on a continuous basis, although there may be missing observations in the data. Markit is one of the most widely employed dataset. Papers that employ this dataset include: Acharya and Johnson (2007), Zhang,

---

[4] The buy-side community includes major credit-focused houses that receive up to 20,000 e-mail pricing messages a day, covering a wide array of credits; and boutique experts focusing on niche credits. These contributors are spread geographically across Europe and the U.S. Each of these members contributes their CDS prices to a CMA database which they receive in Bloomberg formatted messages (as well as forms) from their sell-side dealers. Hence, CMA has access to a constant stream and continuously evolving pool of CDS data. The access to OTC communication between buy-side trading desks and their counterparties guarantees that the prices received by CMA from the buy-side community are very likely to be tradable or even executable prices and that they capture market conditions as they evolve throughout the day. Of course, it is difficult to know precisely whether all of them are tradable or not.

Zhou and Zhu (2009), Jorion and Zhang (2007), Jorion and Zhang (2009), Zhu (2006), Micu et al. (2004), and Cao, Yu, and Zhong (2010).

To summarize, three of the data sources (Reuters EOD, CMA and Markit) employ data from a variety of contributors (over 30 potential dealers/traders) to report composite prices. GFI reports traded CDS spreads. Fenics is a mixture of traded, quoted and calculated CDS spreads all of them based on the same data source and without depending on contributors. CMA, Fenics, GFI and Makit span from January 1, 2004 to March 29, 2010 while Reuters spans from December 3, 2007 to March 29, 2010.

Equity prices are obtained from Reuters. Interest rate related variables employed to capture financing costs of financial institutions and implied volatility indexes are obtained from Datastream, The coefficient of variation of the analysts' forecast of Earnings per Share (EPS) is obtained from I/B/E/S.

For our analysis we use US firms included in the CDX index, as well as European firms included in the iTraxx index. At any point in time, both the CDX and iTraxx indexes contain 125 names each but the composition of the indexes changes every six months. We do not use all the single names CDSs in these indexes but concentrate on the most liquid single names CDSs. As in Christoffersen, Ericsson, Jacobs, and Jin (2009) we use only the single name CDSs which constitute the iTraxx and CDX indexes over the whole sample period which spans from January 2004 to March 2010. We end up with 46 (43) firms which stay in the iTraxx (CDX) index during the whole sample period and for which we are able to obtain equity price and analysts' forecasts information[5].

---

[5] It could be argued that this selection procedure could introduce some survivorship bias in our sample. It should be noted that the components of the indexes are investment grade CDSs firms which are the most actively traded names in the six months prior to the index roll. If in a given period a single name CDS is excluded from the index it is not necessary due to the fact that the firm enters financial distress but simply because of liquidity reasons. On the other hand if a name is downgraded to non-investment it is, of course, excluded from the index. Notice however that one should expect that the agreement among databases on the CDS price for a given name should be higher for the most liquid names. Thus, this possible survivorship

We guarantee a minimum consistency between the single name CDS spread obtained from the different data sources by requiring that all of them have the same maturity (5-year), currency denomination (Euros for the European and US Dollars for the American CDSs), seniority (senior CDS spreads), and restructuring clause (Modified-Modified Restructuring for the European; and Modified Restructuring or No Restructuring, depending on the date, for the American CDS).[6]

Table 1 reports summary statistics on the number of trades and quotes and the mean, standard deviation, and median of the CDS spreads. In Panel A we report the names classified by index and sector for both the American and European firms. Panel B provides aggregated CDS summary statistics for the days in which we have common observations (trades and quotes) in all the data sources distinguishing between European and American firms over the five data sources. The information is divided into two periods: before the crisis (January 1, 2004 – July 31, 2007) and during the crisis (August 1, 2007 – March 2010).[7] Panel C reports the summary statistics for the cases in which there are quotes in all the data sources in a given day independently on whether there are trades or not.[8]

---

bias will tend to make the prices from different databases more in agreement than they are in fact. Consequently if we find significant disagreement among prices from different sources, the empirical evidence is even more compelling.

[6] In the case of Markit, Reuters, Fenics, and GFI; we observe quotes for both the Modified Restructuring and No-Restructuring clauses after April 2009 for the American CDS, coinciding with the Big Bang Protocol. In the case of CMA there is not information on both clauses at the same time but only on the clause for which the contracts are negotiated at that moment. On the basis on this information from CMA, we use the same clause (Modified or No-Restructuring) for the other datasets accordingly.

[7] For each single-name CDS we estimate the optimal number of breakpoints by means of the algorithm described in Bai and Perron (2003) for simultaneous estimation of multiple breakpoints. We focus on the first breakpoint for each CDS comprised between March 2007 and September 2008 and then compute the average breakpoint using the information from all the single-name CDS forming the sample. The average breakpoint corresponds to August 2, 2007 which is in line with the general opinion which sets the beginning of the crisis by August 2007. Looking at concrete events around the former date we find that on July 31, 2007 Bearn Stearns liquidates two hedge funds that invested in various types of mortgage-backed securities. Thus, we decide to consider the last date as the beginning of the crisis period.

[8] We do not report the summary statistics for all the observations given that the existence of missing values in a given date would make more difficult the comparison of the different datasets. In fact, although the crisis period spans from August 2007 to March 2010, the lack of observations from Reuters before December 2007 leads to a crisis period that comprises the dates between December 2007 and March 2010.

For European firms before the crisis, there is price information on trades in 305 days, on average, of the total 870 days in which we have quotes for all the datasets (35% of the days). There is, on average, a considerable degree of agreement among the prices provided by the different databases. The average CDS spread (35 basis points) is similar across the different data sources. However, there are a few firms in which some differences can be found. During the crisis however there is information on transaction prices in 15% of the days in which we have quotes for the remaining four datasets.

Regarding American firms, we observe that before the crisis there is price information on trades in 15% of the days (i.e. 121 days out of the total days in which we have quotes for all the datasets) and there is, on average, a fair amount of agreement among the prices provided by the different databases. However, as is the case with European firms, there are a few names with a relatively similar number of observations where some salient differences can be found. Transaction prices are only available in 2.5% of the days with quotes for the four datasets during the crisis, and the discrepancies are both more frequent and more remarkable.[9,10]

The same pattern in terms of the discrepancies among datasets is observed in Panel C of Table 1: discrepancies are more frequent in the American firms during the crisis. In summary, the preliminary analysis suggests that the crisis has had a strong effect on the degree of disagreement of the different databases in several individual reference entities, and especially so for US names. It is worth noting that, although the total averages are in most cases fairly close, there could be some noteworthy discrepancies both at the entity and also at the cross-sectional level that cannot be captured by these statistics.

---

[9] The low number of observations during the crisis period observed in Panel B is due to the lower number of trades through GFI. The reasons for such a decrease of trades through GFI could include: the effect of the credit crisis that affected the entire CDS market, the market appetite, the competition, or the introduction of the ICE Clearing House among other reasons.

[10] For instance, in the case of American International Group, for a comparable sample with the same dates and equal number of observation, there is a difference of 73 b.p. between the Fenics quote (902 b.p.) and the CMA quote (828 b.p.). Other notable disagreements between the highest and the lowest prices in names in which we compare a sample with the same dates and equal number of observations are found for Comcast (the average difference between Markit and Reuters CDS spreads is more than 70 b.p.), and XL Capital (the average difference between CMA and Fenics is 30 b.p.).

To clarify this point, we depict in Figure 1 the deviations across data sources over time on the basis of two complementary indicators: Root Mean Squared Error (RMSE) in Panels A and B and Relative Root Mean Squared Error (RRMSE) in Panels C and D. The RMSE is obtained as the squared root of the average of the squared of the difference across pairs of data sources[11] (CMA - Markit, CMA - Fenics, CMA - Reuters, Markit - Fenics, Markit - Reuters, Fenics – Reuters) for each firm every day. The RRMSE is obtained as the ratio of the RMSE to the average CDS spread across the four data sources for each firm every day. Panels A and C of Figure 1 report the daily cross-sectional average of the RMSEs and RRMSEs across the total number of firms using the days in which there are trades for the firms. Panels B and D report the cross-sectional averages obtained using the days with no trades. Comparing Panels A and B we observe low RMSEs before the crisis independently on whether there was trading or not. In spite of the low RMSE levels, the average level was significantly larger at any standard level of significance for the days without trades (2.1 b.p. vs. 1.1 b.p.). The average RMSE for the crisis period increases significantly for both series being the increment stronger in the no trading days-firms (from 2.1 b.p. to 11 b.p.) than in the trading days-firms (from 1.1 b.p. to 3.9 b.p.). Of course, the average level for the no trading days (11 b.p.) is significantly larger than the level for the trading days. The first order autocorrelation indicates a stronger persistence of the deviations in the no trading RMSE (0.98) compared to the trading RMSE (0.56). In both series we observe a remarkable increase around Lehman Brother's collapse in September 2008 being the levels of the RMSE larger for the non-trading days-firms. In fact, in Figure B we observe a high dispersion until June 2009.

---

[11] GFI data is not used due to the scarcity of transaction prices during the crisis.

The high levels of dispersion observed in Panels A and B during the crisis could be due to the fact that the levels of credit risks increase during the crisis period. Thus, in Panels C and D we show that the conclusions obtained for Panels A and B also apply when we consider the differences relative to the average level of the CDS spread. The average value of the RRMSE for the series with trades is 3.5% and its volatility is 2%. The average value of the RRMSE for the series without trades is 7% and its volatility is 2.1%. The two sample unpooled t-test with unequal variances has a t-statistic of 32.87 under the null of equal means, suggesting that on average the days without trade information have higher quote dispersion. The same result applies to the pre-crisis and crisis periods. We observe that for the cases with no trades, the average value of the RRMSE is significantly higher during the crisis period relative to the pre-crisis period (6.1% vs. 8.2%). Nevertheless, the levels for the crisis and pre-crisis period are very similar in the cases in which there are trades (3.6% vs. 3.3%). The lower average RRMSE found in the second sub-period suggests that the quotes tend to be more in agreement when there is a trade in a period of financial distress. Thus, the existence of a trade seems to have a stronger effect on the convergence across datasets in the crisis period as it will be confirmed in the later analysis. The RRMSE series obtained using the observations for which there are not transaction prices show a very dynamic behaviour, with some noticeably turbulent episodes. For instance in 2005 given the impact of the crisis experienced by General Motors (GM) and Ford in May 2005 on the credit default swap (CDS) market. The more salient episodes in the RRMSE series are in August 2007 and September 2008. In the days surrounding the Lehman Brothers collapse RRMSE took its highest value to date (15%). Thus, the data suggest that discrepancies from the common trend among databases are persistent and related with market-wide significant episodes. We address the modelling of these discrepancies in Section 4.

Finally, we report the distribution of the differences across data sources in Table 2. This table contains information on the distribution of the quoted and the traded CDS spreads referencing the firms in Table 1. Panel A provides the distribution of the number of quoted spreads on a given day for a single name CDS through the four data sources (CMA, Markit, Reuters and Fenics). Panel B reports the distribution of the range of the mean absolute difference, in basis points, among all the possible pairs of quoted spreads from the previous data sources on a given day for a single CDS. Panel C provides the distribution of the range of the mean absolute difference, in basis points, between all the possible pairs formed by the GFI traded CDS spread and one of the different quoted CDS spreads. In more than 90% of the cases we observe the CDS spread on a given firm at least in three different data sources. The mean absolute difference among the different data sources is higher than one basis point in 55.2% of the cases and it is higher than five basis points in 14.1% of the cases. The mean absolute difference among the traded CDS spread on one hand and the quoted CDS on the other hand (Panel C of Table 2) is slightly higher than the mean differences observed in Panel B of Table 2. Actually the mean difference is smaller than one basis point in 41.9% of the cases.

## 4. Research Hypotheses and Methodology

The main analysis of the data is based on two testable hypotheses. These hypotheses and the methodology employed to perform the empirical tests are detailed in this section.

**Hypothesis 1:** The volatility of the deviations from the common trend of the quoted prices provided by the different CDS data sources is purely random.

In other words, large deviations (in absolute value) from the common trend appear randomly among databases and are unrelated with risk and liquidity factors (global or idiosyncratic). The test of Hypothesis 1 is based on a regression in which the dependent variable is the logarithm of the standard deviation of the 5-year quoted CDS spreads reported by the different data sources which is denoted by $\log(sd(CDS))_{i,t}$ This variable is computed with the $j$ available CDS quoted spreads ($j = 1,..,4$ where $1=$ CMA, $2 =$ Markit, $3 =$ Reuters and $4 =$ Fenics) for a given underlying firm $i$ ($i = 1,..,89$) on every date $t$ as follows: $\log(sd(CDS))_{i,t} = \log((1/n \sum_{j=1}^{n}[CDS_{j,i,t} - (1/n \sum_{j=1}^{n} CDS_{j,i,t})]^2)^{0.5})$, where $n$ is the number of data sources from which we observe CDS spreads, with the maximum $n$ equal four whenever CMA, Markit, Reuters and Fenics report the CDS spreads for firm $i$ at time $t$.

By defining the dependent variable in this way[12] we get rid of the common trend (the average) and concentrate on the deviations from the common trend. The regression equation is as follows:

$$\log(sd(CDS))_{i,t} = \alpha + \beta' X_{k,i,t} + u_{i,t} \quad (1)$$

where the vector $X_{k,i,t}$ includes $k$ explanatory variables: (i) firm's size, which is measured by means of the logarithm of the firm market capitalization, (ii) a trade dummy for the CDS of firm $i$, (iii) number of days without a trade in the CDS of firm $i$, (iv) CDS bid-ask spread, (v) disagreement (coefficient of variation) of the analysts' forecast on the company's earnings per share, (vi) equity price volatility (squared stock returns), (vii) a proxy for the financing costs of financial institutions, and (viii) a proxy for global risk. The vector $\beta'$ includes the regression coefficients corresponding to these $k$ variables while the parameter $\alpha$ is the intercept of the regression. The residual term is denoted by $u_{i,t}$.

---

[12] We take logs to induce the data to meet the assumptions of the regression method that is to be applied; because the distribution of the standard deviation variable is strongly right skewed (the skewness of the original series is 25.10 while the skewness of the log series is 0.21).

The trade dummy is equal to one if there is a trade in the GFI platform at the current date in the 5-year maturity contract, and zero otherwise.[13] The number of days without a trade variable measures the number of days without a trade up to the current date. The proxy for the financing costs of financial institutions is defined as the difference between the 90-day US AA-rated commercial paper interest rate for financial companies and the 90-day US T-bill rate as in Acharya et al. (2007). The proxy for global risk is constructed from the series corresponding to the first principal component of the two main volatility indexes of the two economic areas under study: U.S. (VIX) and Europe (VDAX). The principal component explains 98.5% of the variance of the two previous indexes which confirms the commonalities in terms of global risk between the two indexes. Given the interest in understanding the effect of the constraints on funding for the main participants in the CDS market and the high correlation between the global risk and global liquidity variables (above 0.5); we orthogonalize the two variables by subtracting from the global risk the effect of global financing costs and use the orthogonal residual as the proxy for global risk.[14]

We estimate the coefficients for the above factors using a fixed-effects estimation procedure with standard errors clustered by firm and time (two-way cluster) and robust to heteroskedasticity.[15] Under the null hypothesis, no significant coefficients should be found in equation (1) because differences in price dispersion between databases should be purely random.

---

[13] We are considering trades for the 5-year maturity contract only given that the number of trades in the other maturity contracts is very low. The total number of trades according to GFI information during the sample period and for the firms that we consider is 25,914 while the number of trades which occurred in the other maturities (1 and 3 years contracts) is 1,100 confirming that the most liquid contract is the 5-year CDS contract.

[14] The explanatory variable is the residual of the regression of the series corresponding to the first principal component of the VIX and VDAX indexes onto the proxy for global financing costs and a trend. The residual proxies the level of global risk net of the global financing costs. This strategy also enables us to limit any potential bias derived from collinearity problems.

[15] Hausman's test rejects the random effects specification in favor of a fixed-effects specification, with a p-value of 0.05.

**Hypothesis 2:** The different data sources reflect credit risk information equally efficiently or, equivalently, all databases contribute equally to the price discovery process.

Given that transaction prices are very scarce for some firms, only quoted prices are employed and therefore the comparison is among CMA, Markit, Fenics and Reuters.

To test Hypothesis 2 we employ the Gonzalo and Granger's (1995) model which is based on the following Vector Error Correction Model (VECM) specification and is used to study the effectiveness of the different data sources in terms of price discovery:

$$\Delta X_t = \alpha \beta' X_{t-1} + \sum_{i=1}^{p} \Gamma_i \Delta X_{t-i} + u_t \qquad (2)$$

where equation (2) is formed by a vector autoregressive (VAR) system formed by two equations defined from the vector $X_t$ which includes a pair of CDS quotes or prices of the same underlying firm from two different databases and an error correction term which is defined by the product $\beta' X_{t-1}$ where $\beta' = (1 - \beta_2 - \beta_3)$ are estimated in an auxiliary cointegration regression. The series for the pair of CDS prices included in $X_{t-1}$ must be cointegrated to develop this analysis and the cointegrating relation is defined by $\beta' X_{t-1} = (CDS_{SOURCE\ A, t-1} - \beta_2 - \beta_3 CDS_{SOURCE\ B, t-1})$ which can be interpreted as the long-run equilibrium. The parameter vector α'=(α$_1$ , α$_2$ ) contains the error correction coefficients measuring each price's expected speed in eliminating the price difference and it is the base of the price discovery metrics. The parameter vector $\Gamma_i$ for $i= 1,..p$, with $p$ indicating the total number of lags, contains the coefficients of the VAR system measuring the effect of the lagged first difference in the pair of CDS quotes on the first difference of such quotes at time $t$.[16] Finally, $u_t$ denotes a white noise vector. The percentages of price discovery of the CDS quote $i$ (where $i = 1, 2$) can be defined from the following metrics $GG_i$, $i=1,2$ which are based on the elements of the vector α':

$$GG_1 = \frac{\alpha_2}{-\alpha_1 + \alpha_2}; \qquad GG_2 = \frac{-\alpha_1}{-\alpha_1 + \alpha_2} \qquad (3)$$

The vector α' contains the coefficients that determine each database's contribution to price discovery. Thus, given that $GG_1+GG_2=1$ we conclude that database 1 leads the process of price discovery with respect to database 2 whenever database 1 price discovery metric $GG_1$ is higher than 0.5. If the null hypothesis is true (no dominant market) the percentage of price discovery will be the same for the names from all databases and equal to 0.5. We estimate the price discovery metric for each firm using pairs of CDS spreads and then test whether the average price discovery metric is significantly higher than 0.5 according to the *p-value* that is derived from a bootstrapped *t-statistic*. Each statistic is defined as: $Stat = \frac{(Mean(PDMetrics) - 0.5)}{Std.Dev(PDMetrics)/\sqrt{\#metrics}}$, where # metrics denotes the number of firms for which it is estimated the price discovery metric from a given pair of CDS spreads.

## 5. Empirical Results

### 5.1 Results: Hypothesis 1

#### 5.1.1 Baseline Results

Table 3 shows the regression results obtained from fitting equation (1) to data from the five databases. Column 1 reports the results for the whole sample. Columns 2 and 3 report the results for European and US firms, respectively. Column 4 contains the results for a sample formed by non-financial firms.

---

[16] The optimal number of lags is determined by means of the Schwarz information criteria.

Negative and significant coefficients for the explanatory variable measuring size (logarithm of market capitalization) are found suggesting that the CDS prices for large firms tend to be more in agreement among databases than the prices for small firms. In other words, the volatility of the deviations from the common trend is lower for large firms. This effect is not significant for US firms possibly due to the lower variability of this measure across the US firms. The cross-sectional average volatility of US firm size is around 50% lower than for European firms. This variable is not significant once we exclude the financial firms from the sample which suggests that the deviations between CDS data sources are mainly lower due to the effect of big financial institutions.

The coefficients for the explanatory dummy variable "trade" are negative and significant for all the specifications suggesting that when there are transaction prices available for a given day, the quotes from different contributors tend to agree more closely. The impact of the trades on the deviation across datasets is significantly stronger in the U.S. than in Europe suggesting that the existence of trades lead to a greater convergences of CDS quotes when the number of quotes is scarce. This is in agreement with the results on basic statistical properties summarized in the Section 3 above.

The positive effect found for the variable days w/o trade implies that the longer the period without transaction price information, the greater the disagreement among quotes because, the weaker is the referential value of the previous trade price. One of the reasons explaining the non-significance effect for European firms could be the larger number of trades relative to U.S. CDS contracts.

Regarding the liquidity variable, the bid-ask spread, has, as expected, positive and significant coefficients implying that the more illiquid is the market, the more difficult is to infer appropriate prices and the higher are the deviations from the common trend among the different data sources.

The effect of the proxy for the financing costs or financial constraints of large financial institutions is positive and significant for the four specifications. This result suggests that the shortage of capital held during the crisis by the major financial institutions, which are the main players in the CDS market, affected their activity in this market and so, increased the degree of disagreement among contributors and the dispersion among the contributed CDS prices.

The coefficient of variation of the EPS forecasts of financial analysts enables us to consider the effect of the disagreement in the market regarding the performance of the underlying reference entity. The dispersion of forecasts should capture the uncertainty among analysts due to their private information. According to Batta, Qiu, and Yu (2012) the validity of this variable as a proxy of private information in the CDS market hinges on the similarity between the information environments of CDS investors and financial analysts. Thus, the dispersion in the CDS quotes could be due to the fact that the contributors disagree on the credit quality of the underlying firm according to their private information and so, they quote different prices. In fact, we find that the differences in private information among the analysts have a positive and significant effect on the dispersion of CDS prices.

The use of the proxy for idiosyncratic equity volatility (squared stock returns) enables to control by the underlying firm risk according to the equity market but it could also be a proxy for private information given that the changes in equity prices could reflect private information. The effect of the variable is not significant in any of the four specifications. This non-significant effect could be explained by the limitations of equity volatility as a measure of private information. This variable could misidentify private information in the CDS market unless there is a pooling equilibrium in which the same private information is used for trading in both equity and CDS markets, as pointed out by Batta, Qiu, and Yu (2012).

The proxy for global risk is obtained from the first principal component of the VIX and VDAX indexes. The effect of the volatility index is positive and significant Suggesting that higher global risk implies higher dispersion from the common trend among individual CDS spreads.

One might think that the uncertainty or the disagreement about the credit quality of the financial sector was higher than for the non-financial sectors, especially during the crisis. Nevertheless, the results reported in Column 4, in which we exclude the financial firms from the sample, reject the hypothesis that the dispersion among the CDS databases is mainly driven by the financial sector.[17] Also, model (1) does a pretty good job in explaining the dispersion among prices for the overall sample as measured by the $R^2$ (43%), and also for the European (36%), the US (38%), and the non-financial firms (42%) samples.[18]

The results are quite consistent for all the variables with the exception of size that turns out to be not significant for US and non-financial firms. One potential explanation is that size is only significant on the days without trades. We explore the potential explanation by adding in the baseline regression an interaction term defined as a non-trade dummy multiplied by the firm's size. We obtain that the size variable is still non-significant for the American and non-financial firms but the interaction term is negative and significant for the subgroup of American firms.

---

[17] We have also repeated the analysis for the 19 financial firms in the sample and find similar results.
[18] There should not be any collinearity problem among the explanatory variables given that the maximum correlation among them is the one corresponding to the pair global risk and single-name CDS liquidity and it is equal to 0.37. The low levels of the multi-collinearity condition numbers in the four regressions (the maximum is 4.55 and is well below the critical value 30) confirm that collinearity is unlikely to be a source of concern in our sample.

To summarize, the empirical evidence strongly rejects Hypothesis 1. The volatility of the deviations from the common trend of the quoted prices provided by the different CDS data sources is not random but related to systematic factors. In other words, large deviations from the common trend among databases do not appear randomly but are significantly related to risk, liquidity, trading and firm specific factors. The economic implication of this result is that, in specific market circumstances, the deviations of the prices from the common trend will tend to grow on average. Some prices will be closer to the trend and some prices will be far away from it but the average distance between them will increase, making the prices less homogeneous and making it more difficult for agents to assess the CDS fair value and for researcher using the data to decide what database gives the market prices' most reliable account.

< Insert Table 3 here >

**5.1.2 The drivers of the dispersion of the CDS datasets during the crisis**

Next, we split the analysis using data for the periods before (January 1, 2004 – July 31, 2007) and during the crisis (August 1, 2007 – March 2010), separately. The objective is to check whether the effects of the potential drivers of the dispersion among CDS databases are different for the two above periods. Results are shown in Table 4. Column 1 contains the result for the baseline analysis (Column 1 of Table 3) while Columns 2 and 3 report the results for the pre-crisis and crisis periods, respectively. Column 4 reports the results obtained after including in the baseline analysis a dummy variable for the crisis period that takes value one after August 1, 2007. We observe that in the pre-crisis period all the firm specific factors were significant drivers of the dispersions among data sources. Thus, the larger the disagreement among the analysts' forecast, the larger the volatility in the underlying firms' equity prices, or the smaller the firms; the larger was the dispersion on the CDS prices. The trade dummy also contributes significantly to the CDS prices cross-sectional dispersion while the number of days between two trades does not due, possibly, to the larger number of trades compared to the crisis period. The role of the global variables is more limited being the proxy for financing cost not significant and the proxy for global risk significant at 5%. The explanatory power of the regressors for the pre-crisis period (31.4%) is lower than the one obtained for the crisis period (54.4%) and the whole sample (43.5%).

The results obtained for the crisis period differ from the ones obtained for the pre-crisis period. The variables related to global risk and financing costs are now both highly significant. The strong significant effect of the global factors on the dispersion of the quotes during the crisis could be explained by an increased dependence of default risk on global effects as documented in Aretz and Pope (2011). Among the firm specific variables, the equity volatility is not significant at any standard significance level. The decrease in trading due to the constraints faced by the dealers jointly with the high levels of credit risk observed during the crisis, or the fact that the traders in the equity markets are more heterogeneous and possibly different to the ones of the CDS market; could make this variable a less important indicator of dealers' private information in the CDS market during times of financial distress. The two variables related to the trading activity are significant suggesting a larger effect of the trading on the disagreement of CDS datasets due to the higher effectiveness of transactions in periods with a low market activity. These results suggest a change in the characteristics of the drivers of the lack of agreement among the CDS datasets, being the effect of the trading factors, financing costs, and global risk more significant during than before the crisis. The role of the last two global variables on the dispersion among CDS data sources during the crisis supports Jorion's (2009) evidence about the risk management lessons learned from the crisis. Jorion (2009) points out that the events of 2007 and 2008 highlighted deficiencies and failures of risk models because of known unknowns such as liquidity and counterparty risks.

The Column 4 documents that the crisis contributed to a significant increase in the dispersion across datasets. When the trade variable is included in the regression, the coefficient on trade (-0.085) is lower than the one reported in Column 1 (-0.275), which confirms the higher effectiveness of transactions in the crisis period to diminish the disagreement of CDS datasets. For a more formal test, we include in the regression the interaction between the crisis dummy and the trade variable and find a significant negative effect of this interaction variable confirming the previous statement.

< Insert Table 4 here >

### 5.1.3 Other measures of trading activity

It should be noted that transactions are not necessarily made through the GFI platform, but they could occur in any other platform. The advantage of GFI data is not that it includes all the CDS contracts traded but that it is a transparent source in which the market participants can observe real transaction prices and not just quotes.

Although there is no available data for all the transaction prices since the beginning of our whole sample, we can employ an additional information source for a shorter time period; namely, a "trade information warehouse" that captures the majority of information on CDS trades covering corporate and sovereign borrowers. This warehouse was established by the Depository Trust & Clearing Corporation (DTCC) which keeps a record of outstanding CDSs involving major dealers as counterparties. According to the DTCC calculations around 90-95% of the CDS trades are settled and confirmed through them. The DTCC does not provide all the trade details, which are private information, but it reports weekly data on the gross and net exposures and the number of CDS outstanding contracts on 1,000 corporate and sovereign borrowers. We have this weekly information for the 89 firms that constitute our sample from the 7$^{th}$ of November, 2008 to the last sample date (the 29$^{th}$ of March, 2010).

To test for the importance of trades in the deviations of the CDS prices, we substitute the two trading controls employed in equation (1) by a weekly variable which reports the logarithm of the total number of outstanding CDS contracts traded on a given reference firm. This allows us to control for both the cumulative information on a given firm attending to the total number of contracts and the trend in trading activity. The hypothesis we test is whether a higher number of CDS contacts traded on a given reference firm lowers the volatility of the deviations from the common trend across data sources. Results are shown in Column 2 Table 5. Column 1 contains the results obtained with the baseline specification but using only the trade dummy and observations for the period in which we have information from DTCC (November 2008 – March 2010) for an easier comparison with the other results. We find that the logarithm of the total number of CDS contracts traded on a reference firm has a significant and negative effect on the dispersion between data sources in line with the sign and degree of significance of the trade dummy as observed in Column 1. In fact, the coefficients in Column 2 are similar to the ones of the baseline specification. The implication of this is that the higher the market activity, the lower is the volatility of the deviations from the common trend of the quoted prices provided by the different CDS data sources. This fact is obviously at odds with Hypothesis 1 being true. Additionally, given that we are employing daily information but this variable is constructed on a weekly frequency, we lagged the variable one week and obtain a significant negative coefficient on that variable while the signs and the significance of the other variables remain unchanged.

CMA uses an aggregation methodology which categorizes the prices into "observed" and "derived" depending on the intraday market activity (i.e., the number of buy-side firms receiving prices in a specific risk). Thus, a CDS price is considered as "observed" when CMA receives three different prices from at least two members of is consortium of dealers while the prices that do not fulfil these requirements are considered as "derived". Although these prices are likely to be tradable or even executable prices, it is difficult to know precisely whether all of them are tradable or not. Nevertheless, this variable offers a new opportunity to control by market activity given that the observed prices imply a larger probability for being traded. Thus, we substitute the two trading controls employed in equation (1) by a new proxy that takes value 1 in case the price is observed and 0 in case it is derived. Results are shown in Column 3 of Table 5. This variable has a negative and significant effect on the standard deviation of the quotes of the different CDS data sources confirming the significant role of trading activity. The coefficients for the remaining variables are similar to the ones obtained in the baseline specification (Column 1).

Interestingly, in the three columns of Table 5 none of the firm specific variables with the exception of the CDS bid-ask spread is significant at the 5% level, which confirms the results reported in Table 4 in the sense that during periods of financial distress the firm specific variables lose their explanatory power in favour of the global and trading related variables.

< Insert Table 5 here >

## 5.2 Results: Hypothesis 2

Table 6 reports the results of testing Hypothesis 2 on the price discovery analysis using quoted prices (transaction prices are too scarce to be included in the analysis). A bootstrapped statistical significance test for the null hypothesis, that the estimated price discovery proportions $GG_i$ are equal to 0.5, is also included. The test rejects the null at 5% significance level in all cases with the exception of CMA vs. Markit in Europe and Fenics vs. Reuters also in Europe and in the whole sample. Therefore, in these three cases both databases contribute equally to the price discovery process. However, in all other cases the results indicate that there is a leader database and a follower database. CMA is the data source that contributes to a higher extent to the "formation of prices" with newer and more influential information, especially for the total sample and for the US sample, followed by Markit. For European firms, CMA is slightly more informative than Markit in terms of price discovery. The less informative database in this realm seems to be Fenics, possibly due to the use of the credit-risk model to impute the level of the CDS price when quotes are not available. The results strongly reject the hypothesis that the price discovery process is evenly spread among data bases, and therefore Hypothesis 2 is not supported by the data.

< Insert Table 6 here >

## 6. Impact of CDS data sources discrepancies on price discovery

Given the evidence we present in the previous sections suggesting persistent differences across databases and significant differences in terms of the leadership in the price discovery process, an important question is to what extent these differences translate into conflicting inferences from research based on different databases. Specifically, we focus in one of the most well-known research issues, namely the price discovery process between the stock market and the CDS market. We follow the methodology of Norden and Weber (2009) and direct our attention on the analysis of the price discovery process between stock and CDS returns on the basis of the standard Granger causality test.

Our aim is to assess the degree to which a given database *i* is proximate to database *j* in their price discovery results according to a measure of inconsistency proposed in the network structure literature (see for instance Burt, 1980; or Zuckerman, 2004; among others). Our measure of inconsistency among databases for firm *f* is defined as follows:

$$I_f = \frac{\sum_{i}^{If-1} \sum_{j>i}^{If} i(i,j)}{If(If-1)/2} \quad (4)$$

where *If* is the total number of databases that include firm *f*. The variable *i(i,j)* takes value one when databases *i* and *j* are not in agreement in the price discovery result about firm *f* and is equal to zero otherwise. To be in agreement the equity-CDS price discovery results for the two CDS sources should be in the same of the following categories: (1) CDS returns cause stock returns and stock returns do not cause CDS returns, (2) CDS returns do not cause stock returns and stock returns cause CDS returns, (3) CDS returns cause stock returns and stock returns also cause CDS changes, or (4) CDS returns do not cause stock returns and stock returns do not cause CDS changes.[19] For instance, even if for the two CDS sources we find that the CDS returns cause stock returns but for one of them we find that the stock returns also cause the CDS returns while the stock returns do not cause the CDS returns in the other; we assign value 1 (no agreement). The inconsistency measure varies between one (all databases are not in agreement about firm *f*) and zero (total agreement). We then compute the average inconsistency across all firms and test whether this estimation is statistically different from one and from zero, with the level of significance determined according to the *p-values* that are derived from a bootstrapped *t-statistic*. We also compute the average inconsistency by pairs of data sources on the basis of the four databases containing the most comprehensive records (CMA, Markit, Fenics, and Reuters). The sample period comprises the Lehman Brothers' collapse and spans from March 2008 to April 2009.

---

[19] The level of significance to determine whether stocks cause CDS or vice versa is 1%.

Panel A in Table 7 shows the results of the price discovery test using only firms for which we have information on the stock prices and the four sources of CDS prices for all the days in the sample. Although there is a broad agreement under most scenarios as expected, some cases are worth noting. CMA and Markit give a very similar number of cases across the four potential results or types of causation. Nevertheless, in comparison with CMA and Markit, if a given researcher uses Fenics for a robustness test, she will find more than the double of cases in which stocks cause CDS. If she uses Reuters, she will find around 20% more cases. Furthermore, in comparison with CMA and Markit, the Reuters and Fenics databases finds a much lower number of cases in which stock do not cause CDS. The overall message is that the CMA and Markit results support the hypothesis that in most cases stock returns do not cause CDS returns, but for the other databases the results are less compelling.

Panel B of Table 7 contains the estimation of the average inconsistency across all 89 firms for all the possible pairs of data sources and for all the individual pairs of data sources. The average inconsistency for all firms according to equation (4) is equal to 28%. We use two significance tests to check whether this figure represents a significant inconsistency or not. The null hypotheses in the two tests are: (1) full inconsistency and (2) full consistency. In both cases the data clearly reject the null hypotheses suggesting that on average all databases are neither in full disagreement nor in total agreement about the price discovery results for a given firm $f,$ although the average inconsistency measure is closer to zero than to one. The implication is therefore that although significant inconsistency is found across databases, the price discovery results tend to be in agreement more than in disagreement. We next compare the consistency or inconsistency of the data sources by pairs. In line with the results presented in Panel A, we reject the existence of a total consistency in the different pairs of data sources but we also reject the total inconsistency. According to these results, we conclude that the use of Markit or CMA would lead to similar results. In fact the inconsistency between these two data sources is not significant at 1% and they only disagree in around 10% of the cases. Nevertheless, the inconsistency of Markit or CMA and the other two data sources is remarkable (between 24% and 38% of the cases exhibit disagreement) as it is also the case for the inconsistency measure between Fenics and Reuters.

< Insert Table 7 here >

## 7. Robustness Tests

In this section, we report the results of several checks of the test of Hypothesis 1 presented in Table 3.

**7.1. Alternative combinations of data sources to measures their dispersion**

All the data sources, with the exception of Fenics, are based on traders or dealers prices. As was mentioned in Section 3, Fenics data can be actual trades or mid-prices calculated from the bid/offer quotes. If none of these are available, GFI, which is the responsible of the Fenics quotes, calculates the CDS spread using the Hull and White methodology to ensure that a credit curve always exists for each reference entity. Bedendo, Cathcart and El-Jahel (2011) find that the gap between CDS model spreads and CDS market spreads is time varying and widens substantially in times of financial turbulence partly due to the shortcomings of the model. Thus, our aim is to check whether the effects of the explanatory variables on the differences in the CDS quotes are due to the use of the Fenics CDS quotes obtained from credit risk models that use asset volatility and market volatility as inputs. To check the last statement, we repeat regression (1) using as the dependent variable the logarithm of the standard deviation using only CMA, Markit, and Reuters EOD. These three datasets are more homogenous since they collect data from nearly 30 contributors. Results are shown in Column 2 of Table 8 (Column 1 contains the baseline results). The coefficients are similar to the ones for the baseline dependent variable. The coefficients for the trading dummies have a lower impact than the ones obtained including Fenics but they are still significant. The effect of size is significantly larger when we exclude Fenics possibly because Fenics credit-risk model does not consider size as a relevant input to obtain the price of credit risk.

The data obtained from Reuters EOD is available from December 2007, whereas the remaining data sources have information starting from January 2004. To avoid any potential bias due to the different length of the sample period covered by the different data sources we repeat the previous analysis without including the Reuters EOD quotes. Results are reported in Column 3 of Table 8. The new results are again almost identical to the ones obtained in the baseline analysis confirming the robustness of our results.

< Insert Table 8 here >

**7.2 Other econometric methodologies and data frequencies**

We first check for the robustness of our results to pure time effects. Instead of using a full set of time dummies (1,600 day dummies in our analysis); we calculate the mean of the dependent variable on a daily basis and subtract this daily mean for each firm CDS day by day. Time fixed effects control for all factors that are invariant across firms in a given time period, including global risk and financing costs. Thus, we exclude from this estimation global risk and financing costs variables. The results are reported in Column 2 of Table 9 (Column 1 contains the baseline results reported in Column 1 of Table 3). We conclude that the results are robust to this treatment.[20]

The quotes contributed to the different data sources employed in our paper are gathered over the entire trading day from London open to New York close. These quotes are aggregated by the data providers to report a daily (end-of-day) quote after the close of the New York market.[21] That is, our data does not rely on intraday CDS prices updates but on the update that occurs after the close of the New York market. Of course, the exact times of the different quotes employed by the data providers to compute the daily quote do not necessarily refer exactly to the same times and it will depend on the times at which the contributors are willing to quote a price or trade a contract.

To avoid timing pitfalls and mitigate any potential bias induced by infrequent and/or nonsynchronous quotes we repeat the baseline analysis using data with a lower frequency as suggested by Harvey (1995). We estimate equation (1) using weekly instead of daily data. Results, reported in Column 3 of Table 9, confirm the robustness of our analysis and the fact that the non-synchronization among data sources is not a serious issue in this study.[22]

< Insert Table 9 here >

### 7.3. Alternative dependent variable and sample periods

---

[20] Given that the data is unbalanced we use as an alternative econometric technique Prais-Winsten regressions with correlated panels, corrected standard errors (PCSEs) and robust to heteroskedasticity, contemporaneous correlation across panels and serial autocorrelation within panels and find similar results to the ones reported in this paper. The results obtained under this alternative methodology are available upon request.

[21] The time limit for the dealers to contribute the prices varies across the data providers. For instance, the limit imposed to the dealers by Markit and Reuters for submitting the prices is 01.00 AM GMT (the price is made publicly available some hours later. CMA and Fenics provide the price just after the New York market close given that CMA gets the quotes directly from the communications between the largest counterparties and Fenics from their own trading desks.

[22] The variable referred to the number of weeks without quotes is not included in this analysis given that for many firms it takes value 0 for the whole sample period, especially before the crisis and for the European firms. The dummy referred to the existence of trades in GFI for a given day is now transformed into a variable that takes values from 0 to 5 indicating the number of days with trades within a given week.

The dependent variable that we employed in the previous analysis is defined in logs in order to limit the effect of potential outliers which could appear in the quoted spreads due to any mistake in the contributed prices. By using the logs we also limit potential problems derived from a skewed distribution given that the value of the mean is almost four times the value of the median. We repeat regression (1) using as the dependent variable the ratio between the logarithm of the standard deviation among the CDS quotes and the logarithm of mean across the CDS quotes. The results are almost identical to the ones reported in Table 3.

Since the beginning of the financial crisis counterparty risk in the CDS contracts has been partially mitigated through the use of collateralization. Actually, full collateralization of CDS liabilities has become the market standard. The ISDA Margin Survey 2009 reports that 74% of CDS contracts executed during 2008 were subject to collateral agreements. In order to limit any potential difference in the use of this collateral by the CDS data source we repeat the same analysis using a sub sample which spans up to December 2007 given that the use of the collateral was more limited before 2008. The results do not materially differ from the ones reported in the first column of Table 3. The only significant difference is that the coefficient of the variable related to the number of days without quote, although with a positive sign, is not significant now (p-value = 0.12). Moreover, the squared stock returns are now significant at 1% level in line with the result that suggests that the idiosyncratic volatility has a significant effect in more tranquil periods.

On April 8, 2009, ISDA introduced the "Big Bang Protocol" that aims to introduce more consistency into the CDS market by imposing a uniform procedure for settling CDS contracts when a company goes into default. Additionally, this Protocol tries to impose more standardization by introducing set coupons for contracts. This measure was initially limited to the US but later spread into Europe. To discard any potential effect on CDS quotes derived from the introduction of the Big Bang Protocol, mainly in the case of the American CDS, we repeat the same analysis using data until April 8, 2009.[23] The results are similar to the ones obtained using the baseline specification both in terms of the sign and the degree of significance of the coefficients.

The results obtained in the different robustness tests in this subsection are not reported for the sake of brevity but are available upon request.

## 8. Conclusions

We study the degree of likeness of the five most widely used CDS data bases: GFI, Fenics, Reuters EOD, CMA, and Markit, for the period from 2004 to 2010 using the most liquid single name 5-year CDS of the components of the leading market indexes, iTraxx (European firms) and CDX (US firms). We find that there are significant differences among them in several dimensions.

Our main empirical findings are:

1)      When timely information on traded prices is available, the different price sources largely agree among them. However, as the information on transaction prices become scarcer, prices from different sources tend to diverge from the common trend.  The most extreme disagreements are in the case of American reference entities during the crisis, for which very few transaction prices are available in the GFI database.

2)      Deviations from the common trend among the different CDS quoted spreads are not purely random but are related to idiosyncratic factors such as firm size, equity prices volatility, and disagreement of analysts about EPS of the firm; and also to single-names CDS liquidity, financing costs, global risk, and trading factors. Prices for a given CDS contract tend to diverge less from the common trend in the case of bigger firms and more frequently traded contracts. Increases in CDS illiquidity, disagreement about EPS of the underlying firms, financing costs for financial institutions, and global implied volatility; increase the divergence from the common trend among prices coming from different data bases.

---

[23] One of the key developments in restoring market confidence was Intercontinental Exchange's (ICE) introduction of CDS clearing in March 2009. We have repeated the analysis using data until March 2009 and find similar results. Looking forward, we aim to test what are the implications of the central clearing on the issues raised in this paper and to verify whether we still need to worry about data quality issued from different sources.

3) CMA quoted CDS spreads led the credit risk price discovery process with respect to the quotes provided by the other databases.

4) There is not a full consistency among the databases in the results of price discovery analysis between stock and CDS returns.

Extensions and robustness tests support these results. Since our analysis is based on the most liquid CDS prices, we would expect that the differences we find for these prices in the different databases would be even larger for less liquid CDS not included in our study.

The most important implications for research studies and industry participants that arise from this study are twofold. First, in times of high illiquidity, high financing costs, and high global risk, CDS prices from different databases will tend to substantially diverge from the common trend. This makes more difficult for agents to disentangle the CDS fair value from the prices of the different databases and for researchers to decide what database gives the market prices' most reliable account. Second, extant price discovery research focused on CDS spreads and/or stock returns should be complemented with additional robustness tests in the light of the differences found across databases.

Avenues of further research include the analysis of the effect of the discrepancies among the different CDS contributors on potential debt decoupling (Hu and Black, 2008) or the usefulness of such dispersion as a signaling tool for supervisory monitoring in comparison to the usefulness of other financial instruments (Hamalainen, Pop, Hall and Howcroft, 2012).

Zuckerman, E.W., "Structural Incoherence and Stock market activity". *American Sociological Review*, Vol. 69, 2004, pp.405-432

# Table 1: Firm Names by Sector and CDS Index (iTraxx and CDX)

This table shows the descriptive statistics for the single name 5-year CDS. Panel A shows the names classified by index and sector. We use European and American firms included in the iTraxx and the CDX indexes, respectively, over the whole sample period. Panel B provides aggregated CDS summary statistics for the five data sources (GFI, CMA, Markit, Fenics, and Reuters) using the days in which we have common observations (trades and quotes) in all the data sources distinguishing between European and American firms. The information is divided into two periods: before (January 2004 – July 2007) and during (August 2007 – March 2010) the financial crisis. Panel C reports the summary statistics for the cases in which there are quotes in four data sources (CMA, Markit, Fenics, and Reuters) in a given day independently on whether there are trades or not

Panel A

| iTraxx Firm Name | Ticker | Sector | CDX Firm Name | Ticker | Sector |
|---|---|---|---|---|---|
| AKZO Nobel NV | AKZO | Auto/Indust. | Alcoa Inc. | AA | Auto/Indust. |
| Bayer Aktiengesellschaft | BAYG | Auto/Indust. | Carnival Corporation | CCL | Auto/Indust. |
| Bayerische Motoren Werke AG | BMWG | Auto/Indust. | CSX Corporation | CSX | Auto/Indust. |
| Compagnie de Saint-Gobain | SGOB | Auto/Indust. | The Dow Chemical Company | DOW | Auto/Indust. |
| EADS NV | AERM | Auto/Indust. | Eastman Chemical Company | EMN | Auto/Indust. |
| Siemens Aktiengesellschaft | SIEG | Auto/Indust. | Honeywell International Inc | HON | Auto/Indust. |
| Volkswagen Aktiengesellschaft | VOWG | Auto/Indust. | Union Pacific Corporation | UNP | Auto/Indust. |
| Aktiebolaget Volvo | VOLV | Auto/Indust. | | | |
| Accor | ACCP | Consumers | Altria Group, Inc. | MO | Consumers |
| British American Tobacco PLC | BATS | Consumers | AutoZone, Inc. | AZO | Consumers |
| Carrefour | CARR | Consumers | Baxter International Inc. | BAX | Consumers |
| Marks and Spencer PLC | MKSA | Consumers | Bristol-Myers Squibb Company | BMY | Consumers |
| LVMH Moet Hennessy Louis Vuitton | LVMH | Consumers | Campbell Soup Company | CPB | Consumers |
| Metro AG | METB | Consumers | Cardinal Health, Inc. | CAH | Consumers |
| Koninklijke Philips Electronics NV | PHG | Consumers | Loews Corporation | LTR | Consumers |
| PPR | PRTP | Consumers | Safeway Inc. | SWY | Consumers |
| Sodexho Alliance | SODE | Consumers | Southwest Airlines Co. | LUV | Consumers |
| Unilever NV | UN | Consumers | The Walt Disney Company | DIS | Consumers |
| | | | Whirlpool Corporation | WHR | Consumers |
| Edison SPA | EDN | Energy | Anadarko Petroleum Corporation | APC | Energy |
| Electricite de France | EDF | Energy | Arrow Electronics, Inc. | ARW | Energy |
| EnBW Energie Baden-Wuerttemberg | EBKG | Energy | ConocoPhillips | COP | Energy |
| Enel SPA | ENEI | Energy | Constellation Energy Group, Inc. | CEG | Energy |
| EDP - Energias de Portugal SA | EDP | Energy | Devon Energy Corporation | DVN | Energy |
| E.ON AG | EONG | Energy | Dominion Resources, Inc. | D | Energy |
| Fortum Oyj | FUMC | Energy | Progress Energy, Inc. | PGN | Energy |
| Iberdrola SA | IBE | Energy | Sempra Energy | SRE | Energy |
| Repsol YPF SA | REP | Energy | Transocean Inc. | RIG | Energy |
| RWE Aktiengesellschaft | RWEG | Energy | Valero Energy Corporation | VLO | Energy |
| GDF Suez | GDF | Energy | | | |
| Veolia Environnement | VIE | Energy | | | |
| Aegon NV | AEGN | Financials | Ace Limited | ACE | Financials |
| AXA | AXAF | Financials | American Express Company | AXP | Financials |
| Barclays Bank PLC | BCSB | Financials | American International Group, Inc. | AIG | Financials |
| Commerzbank Aktiengesellschaft | CBKG | Financials | Boeing Capital Corporation | BA | Financials |
| Deutsche Bank Aktiengesellschaft | DB | Financials | Cigna Corporation | CI | Financials |
| Hannover Rueckversicherung AG | HNRG | Financials | General Electric Capital Corporation | GE | Financials |
| Banca Monte Dei Paschi Di Siena Spa | BMPS | Financials | Marsh & McLennan, Inc. | MMC | Financials |
| Muenchener Rueckversicherung | MUVG | Financials | Simon Property Group, L.P. | SPG | Financials |
| Swiss Reinsurance Company | RUKN | Financials | Wells Fargo & Company | WFC | Financials |
| | | | XL Capital Ltd. | XL | Financials |
| Deutsche Telekom AG | DTA | TMT | AT&T Inc. | T | TMT |
| France Telecom | FTE | TMT | CenturyTel, Inc. | CTL | TMT |
| Hellenic Telecommunications | OTE | TMT | Comcast Cable Communications, LLC | CMCC | TMT |
| Koninklijke KPN NV | KPN | TMT | Omnicom Group Inc. | OMC | TMT |
| Telecom Italia SPA | TLIT | TMT | Time Warner Inc. | TWX | TMT |
| Telefonica SA | TEF | TMT | | | |
| Vodafone Group PLC | VOD | TMT | | | |

**Panel B: Using the observations in the days in which there is a trade and quotes in all the data sources**

| Europe | Average Number of Trades and Quotes | Mean | S.D. | Median | Average Number of Trades and Quotes | Mean | S.D. | Median |
|---|---|---|---|---|---|---|---|---|
| | Before 31st July 2007 | | | | After 1st August 2007 | | | |
| GFI | 305 | 36 | 17 | 33 | 72 | 118 | 74 | 99 |
| CMA | 305 | 35 | 16 | 32 | 72 | 118 | 74 | 99 |
| Markit | 305 | 35 | 16 | 31 | 72 | 118 | 74 | 98 |
| Fenics | 305 | 35 | 16 | 32 | 72 | 118 | 74 | 99 |
| Reuters | | | | | 72 | 117 | 74 | 98 |

| US | Average Number of Trades and Quotes | Mean | S.D. | Median | Average Number of Trades and Quotes | Mean | S.D. | Median |
|---|---|---|---|---|---|---|---|---|
| GFI | 121 | 54 | 36 | 44 | 12 | 130 | 120 | 100 |
| CMA | 121 | 52 | 36 | 44 | 12 | 150 | 130 | 113 |
| Markit | 121 | 52 | 36 | 44 | 12 | 150 | 129 | 114 |
| Fenics | 121 | 52 | 37 | 44 | 12 | 134 | 122 | 97 |
| Reuters | | | | | 12 | 147 | 121 | 114 |

**Panel C: Using the observations in the days in which there are quotes in all the data sources**

| Europe | Average Number Quotes | Mean | S.D. | Median | Average Number Quotes | Mean | S.D. | Median |
|---|---|---|---|---|---|---|---|---|
| | Before 31st July 2007 | | | | After 1st August 2007 | | | |
| CMA | 854 | 31 | 18 | 27 | 504 | 117 | 93 | 89 |
| Markit | 854 | 31 | 18 | 27 | 504 | 117 | 93 | 89 |
| Fenics | 854 | 31 | 18 | 27 | 504 | 118 | 94 | 89 |
| Reuters | | | | | 504 | 117 | 93 | 89 |

| US | Average Number Quotes | Mean | S.D. | Median | Average Number Quotes | Mean | S.D. | Median |
|---|---|---|---|---|---|---|---|---|
| CMA | 735 | 38 | 24 | 34 | 490 | 138 | 202 | 85 |
| Markit | 735 | 38 | 24 | 34 | 490 | 140 | 202 | 86 |
| Fenics | 735 | 39 | 24 | 35 | 490 | 131 | 216 | 82 |
| Reuters | | | | | 490 | 136 | 206 | 82 |

**Table 2: The distribution of the quoted and traded CDS spreads**

Summary statistics for the distribution of the quoted and traded 5y CDS spreads for the firms in Table 1. Panel A provides the distribution of the number of quoted spreads on a given day for a single name CDS. The first column reports the number of observations for which there are 1, 2, 3 and 4 quotes, respectively. The second column reports the percentage of cases in which there are 1, 2, 3, and 4 quotes, respectively, and is obtained as the ratio of the first column and the total number of days. The last column of Panel A reports the cumulative percentage across the range of quotes. Panel B reports the distribution of the range of the mean absolute difference, in basis points, among all the possible pairs of quoted spreads on a given day for a single 5-year CDS. For instance, the first row in the first column reports the number of days in which the absolute difference between the different pairs of databases is lower than 1 basis point, and so on. Panel C provides the distribution of the range of the mean absolute difference, in basis points, between all the possible pairs formed by the GFI traded CDS spread and one of the different quoted CDS spreads. The sample period is January 2004 to March 2010.

Panel A

| Number of Quotes | Number of Observations | Percentage | Cumulative Percentage |
|---|---|---|---|
| 1 | 1,235 | 0.009 | 0.009 |
| 2 | 11,968 | 0.087 | 0.096 |
| 3 | 81,246 | 0.588 | 0.683 |
| 4 | 43,767 | 0.317 | 1.000 |
| Total Days | 138,216 | | |
| Total Trades | 25,914 | 0.187 | |

Panel B

| Range of the mean difference among quotes (in b.p.) | Observations | Percentage | Cumulative Percentage |
|---|---|---|---|
| Dif < 1 | 61,319 | 0.448 | 0.448 |
| 1 < Dif < 2 | 27,490 | 0.201 | 0.648 |
| 2 < Dif < 3 | 12,770 | 0.093 | 0.742 |
| 3 < Dif < 4 | 7,365 | 0.054 | 0.795 |
| 4 < Dif < 5 | 5,100 | 0.037 | 0.833 |
| 5 < Dif < 6 | 3,613 | 0.026 | 0.859 |
| 6 < Dif < 7 | 3,007 | 0.022 | 0.881 |
| 7 < Dif < 8 | 2,256 | 0.016 | 0.897 |
| 8 < Dif < 9 | 1,814 | 0.013 | 0.911 |
| 9 < Dif < 10 | 1,445 | 0.011 | 0.921 |
| 10 < Dif < 11 | 1,234 | 0.009 | 0.930 |
| 11 < Dif < 12 | 1,066 | 0.008 | 0.938 |
| 12 < Dif < 13 | 874 | 0.006 | 0.944 |
| 13 < Dif < 14 | 722 | 0.005 | 0.950 |
| 14 < Dif < 15 | 692 | 0.005 | 0.955 |
| 15 < Dif < 16 | 489 | 0.004 | 0.958 |
| 16 < Dif < 17 | 441 | 0.003 | 0.961 |
| 17 < Dif < 18 | 348 | 0.003 | 0.964 |
| 18 < Dif < 19 | 340 | 0.002 | 0.966 |
| 19 < Dif < 20 | 300 | 0.002 | 0.969 |
| Dif >= 20 | 4,296 | 0.031 | 1.000 |
| Total Quotes | 136,981 | | |
| Missing Obs. (just 1 quote) | 1,235 | | |
| Total Quotes | 138,216 | | |

Panel C

| Range of the mean difference between the traded and quoted spreads (in b.p.) | Observations | Percentage | Cumulative Percentage |
|---|---|---|---|
| Dif < 1 | 10,852 | 0.419 | 0.419 |
| 1 < Dif < 2 | 5,310 | 0.205 | 0.624 |
| 2 < Dif < 3 | 2,539 | 0.098 | 0.722 |
| 3 < Dif < 4 | 1,640 | 0.063 | 0.785 |
| 4 < Dif < 5 | 1,222 | 0.047 | 0.832 |
| 5 < Dif < 6 | 837 | 0.032 | 0.864 |
| 6 < Dif < 7 | 627 | 0.024 | 0.889 |
| 7 < Dif < 8 | 407 | 0.016 | 0.904 |
| 8 < Dif < 9 | 299 | 0.012 | 0.916 |
| 9 < Dif < 10 | 266 | 0.010 | 0.926 |
| 10 < Dif < 11 | 242 | 0.009 | 0.935 |
| 11 < Dif < 12 | 174 | 0.007 | 0.942 |
| 12 < Dif < 13 | 241 | 0.009 | 0.951 |
| 13 < Dif < 14 | 167 | 0.006 | 0.958 |
| 14 < Dif < 15 | 116 | 0.004 | 0.962 |
| 15 < Dif < 16 | 104 | 0.004 | 0.966 |
| 16 < Dif < 17 | 90 | 0.003 | 0.970 |
| 17 < Dif < 18 | 52 | 0.002 | 0.972 |
| 18 < Dif < 19 | 39 | 0.002 | 0.973 |
| 19 < Dif < 20 | 22 | 0.001 | 0.974 |
| Dif >= 20 | 668 | 0.026 | 1.000 |
| Total Quotes | 25,914 | | |

**Table 3: Determinants of the deviation
among the CDS data sources**

This table reports the regression coefficients for the potential drivers of the dispersion among data sources. The dependent variable is the logarithm of the standard deviation among the different CDS data sources (CMA, Markit, Fenics, and Reuters EOD). The database includes 89 European and US firms (46 of the firms are European and 43 are American) which are the most liquid CDSs included in either the Itraxx or the CDX Index since the launching of the indexes, from January 2004 to March 2010. We estimate the coefficients for the above factors using a fixed-effects estimation procedure with standard errors clustered by firm and time (two-way cluster) and robust to heteroskedasticity. Column 1 reports the results for the whole sample of firms; Columns 2 and 3 report the results for the subsample of European and American firms, respectively; and Column 4 reports the results for the subsample of non-financial firms. Each column contains the explanatory variables' coefficients and the standard errors between brackets. *** (** and *) indicates whether the coefficients are significant at a significance level of 1% (5% and 10%).

|  | (1) | (2) | (3) | (4) |
|---|---|---|---|---|
| Size (logarithm of market cap.) | -0.181** | -0.229** | -0.180 | -0.099 |
|  | (0.09) | (0.11) | (0.14) | (0.09) |
| Trade | -0.275*** | -0.184*** | -0.410*** | -0.233*** |
|  | (0.04) | (0.02) | (0.09) | (0.04) |
| Days w/o a trade | 0.001*** | 0.0000 | 0.001*** | 0.001*** |
|  | (0.00) | (0.00) | (0.00) | (0.00) |
| CDS Bid-Ask Spread | 0.054*** | 0.097*** | 0.040** | 0.105*** |
|  | (0.02) | (0.01) | (0.02) | (0.01) |
| Coefficient of variation of analysts' forecasts of EPS | 0.0002*** | 0.0001*** | 0.002*** | 0.0001*** |
|  | (0.00) | (0.00) | (0.00) | (0.00) |
| Equity volatility (squared stock returns) | -0.645 | 0.777 | -3.218 | 0.594 |
|  | (0.86) | (0.65) | (3.93) | (0.84) |
| Financing costs of financial institutions | 0.496*** | 0.400*** | 0.526*** | 0.409*** |
|  | (0.05) | (0.03) | (0.06) | (0.03) |
| Global risk (common implied volatility) | 0.025*** | 0.013*** | 0.029*** | 0.017*** |
|  | (0.00) | (0.00) | (0.00) | (0.00) |
| Constant | 3.903* | 4.467* | 4.317 | 1.657 |
|  | (2.21) | 2.62 | (3.47) | (2.19) |
| R-squared | 0.430 | 0.360 | 0.380 | 0.420 |
| Number of observations | 132550 | 67534 | 65016 | 104698 |
| Number of groups | 89 | 46 | 43 | 70 |
| F-statistic | 4228.370 | 2761.190 | 2114.220 | 3733.910 |
| Prob. > F-statistic | 0 | 0 | 0 | 0 |
| Condition Index | 3.53 | 4.27 | 3.45 | 4.55 |

**Table 4: Determinants of the deviation
among the CDS data sources before and during the crisis**

This table reports the regression coefficients for the potential drivers of the dispersion among data sources before and during the subprime crisis. The dependent variable is the logarithm of the standard deviation among the different CDS data sources (CMA, Markit, Fenics, and Reuters EOD). The database includes 89 European and US firms (46 of the firms are European and 43 are American) which are the most liquid CDSs included in either the Itraxx or the CDX Index since the launching of the indexes, from January 2004 to March 2010. We estimate the coefficients for the above factors using a fixed-effects estimation procedure with standard errors clustered by firm and time (two-way cluster) and robust to heteroskedasticity. Column 1 reports the results for the whole sample of firms and is equivalent to the baseline regression estimation (Column 1 of Table 3); Column 2 reports the estimation results using information before July 31, 2007, which is set as the date for the beginning of the subprime crisis; Column 3 contains the estimation results using information during the subprime crisis period (after July 31, 2007). Column 4 reports the results obtained after including in the baseline regression a dummy variable for the crisis period that takes value one after August 1, 2007. Each column contains the explanatory variables' coefficients and the standard errors between brackets. *** (** and *) indicates whether the coefficients are significant at a significance level of 1% (5% and 10%).

|  | (1) | (2) | (3) | (4) |
|---|---|---|---|---|
| Size (logarithm of market cap.) | -0.181** | -0.343*** | -0.231** | -0.207*** |
|  | (0.09) | (0.11) | (0.09) | (0.05) |
| Trade | -0.275*** | -0.054** | -0.135*** | -0.085*** |
|  | (0.04) | (0.03) | (0.03) | (0.03) |
| Days w/o a trade | 0.0008*** | 0.0002 | 0.0003** | 0.0005*** |
|  | (0.00) | (0.00) | (0.00) | (0.00) |
| CDS Bid-Ask Spread | 0.054*** | 0.042*** | 0.030** | 0.037*** |
|  | (0.02) | (0.01) | (0.01) | (0.01) |
| Coefficient of variation of analysts' forecasts of EPS | 0.0002*** | 0.0000*** | 0.0001*** | 0.0001*** |
|  | (0.00) | (0.00) | (0.00) | (0.00) |
| Equity volatility (squared stock returns) | -0.645 | 42.058*** | 0.063 | 0.691 |
|  | (0.86) | (9.40) | (0.45) | (0.48) |
| Financing costs of financial institutions | 0.496*** | 0.131 | 0.122*** | 0.119*** |
|  | (0.05) | (0.15) | (0.03) | (0.03) |
| Global risk (common implied volatility) | 0.025*** | 0.010** | 0.023*** | 0.020*** |
|  | (0.00) | (0.00) | (0.00) | (0.00) |
| Crisis dummy |  |  |  | 1.005*** |
|  |  |  |  | (0.08) |
| Constant | 3.903* | 7.578*** | 6.012*** | 4.356*** |
|  | (2.21) | (2.63) | (2.17) | (1.29) |
| R-squared | 0.43 | 0.31 | 0.54 | 0.51 |
| Number of observations | 132550 | 75288 | 57262 | 132550 |
| Number of groups | 89 | 89 | 89 | 89 |
| F-statistic | 4228.37 | 203.33 | 1846.57 | 7845.63 |
| Prob. > F-statistic | 0 | 0 | 0 | 0 |
| Condition Index | 3.53 | 8.31 | 3.69 | 4.25 |

**Table 5: The role of trading activity as a determinant of the deviation among the CDS data sources**

This table reports the regression coefficients for the potential drivers of the dispersion among CDS data sources using three different proxies for the trading activity. The dependent variable is the logarithm of the standard deviation among the different sources (CMA, Markit, Fenics, and Reuters EOD). The database includes 89 European and US firms (46 European and 43 American firms) which are the most liquid CDSs included in either the Itraxx or the CDX Index since the launching of the indexes to March 2010. The coefficients for the above are estimated on the basis of a fixed-effects estimation procedure with standard errors clustered by firm and time (two-way cluster) and robust to heteroskedasticity. The sample spans from November 2008 to March 2010 due to the lack of information before this date for the trading activity proxy that is constructed from DTCC data. This sample length leads to an easier comparison of the results corresponding to the three proxies for trading activity. Column 1 reports the results for the case in which we the proxy takes value 1 in case there is a trade in GFI (the same employed in the baseline specification). Column 2 reports the results obtained when the proxy for trading activity is the logarithm of the total number of outstanding CDS contracts traded on a given reference firm that is obtained from DTCC. The trading activity proxy in Column 3 is an indicator that takes value 1 in case the daily quotes reported by CMA are observed and 0 in case they derived. A quote is considered as observed when CMA receives three different prices from at least two dealers; otherwise it is derived. Each column contains the explanatory variables' coefficients and the standard errors between brackets. *** (** and *) indicates whether the coefficients are significant at a significance level of 1% (5% and 10%).

|  | (1) | (2) | (3) |
|---|---|---|---|
| Size (logarithm of market cap.) | -0.138 | -0.156 | -0.144 |
|  | (0.14) | (0.13) | (0.14) |
| Trade | -0.103** |  |  |
|  | (0.04) |  |  |
| Logarithm of the number of CDS contracts traded |  | -0.369** |  |
|  |  | (0.16) |  |
| Observed/derived CDS quote indicator |  |  | -0.295*** |
|  |  |  | (0.07) |
| CDS Bid-Ask Spread | 0.043*** | 0.042*** | 0.042*** |
|  | (0.01) | (0.01) | (0.01) |
| Coefficient of variation of analysts' forecasts of EPS | 0.0001* | 0.0001* | 0.0001* |
|  | (0.00) | (0.00) | (0.00) |
| Equity volatility (squared stock returns) | -0.882 | -0.821 | -0.608 |
|  | (2.02) | (1.98) | (2.02) |
| Financing costs of financial institutions | 0.115*** | 0.116*** | 0.106*** |
|  | (0.04) | (0.04) | (0.04) |
| Global risk (common implied volatility) | 0.016*** | 0.015*** | 0.016*** |
|  | (0.00) | (0.00) | (0.00) |
| Constant | 3.833 | 7.247** | 4.242 |
|  | (3.38) | (3.42) | 3.34 |
| R-squared | 0.623 | 0.624 | 0.624 |
| Number of observations | 28867 | 28867 | 28867 |
| Number of groups | 88 | 88 | 88 |
| F-statistic | 983.750 | 1007.430 | 1047.290 |
| Prob. > F-statistic | 0 | 0 | 0 |
| Condition Index | 3.82 | 6.82 | 12.50 |

**Table 6: Price Discovery Analysis by Pairs of CDS spreads**

This table reports the results of the price discovery analysis. First, we estimate the Gonzalo-Granger (GG) price discovery metrics for different pairs of 5-year single name CDS spreads using different data sources. The estimations are based on a VECM in which the VAR-length is selected according to the Schwarz information criteria. Then we calculate the average Gonzalo-Granger metric for all the firms, the European, and the American firms for the different pairs of data sources. When the price discovery metric is higher than 0.5, the corresponding data source leads the price discovery process. The symbols ***, **, and * (^^^, ^^, and ^) summarize the statistical significance test and indicate that the average price discovery metric (GG) corresponding to a given data source is significantly higher (lower) than 0.5 at a significance level of 1, 5 and 10%, respectively. The level of significance is determined according to *p-values* that are derived from a bootstrapped *t-statistic*.

CMA versus Markit GG Price Discovery Metrics

|        | Total    | Europe  | US       |
|--------|----------|---------|----------|
| CMA    | 0.63**   | 0.59*   | 0.67***  |
| Markit | 0.37^^   | 0.41^   | 0.33^^^  |

CMA versus Fenics GG Price Discovery Metrics

|        | Total    | Europe   | US       |
|--------|----------|----------|----------|
| CMA    | 0.75***  | 0.77***  | 0.72***  |
| Fenics | 0.25^^^  | 0.23^^^  | 0.28^^^  |

CMA versus Reuters EOD GG Price Discovery Metrics

|             | Total    | Europe   | US       |
|-------------|----------|----------|----------|
| CMA         | 0.79***  | 0.83***  | 0.74***  |
| Reuters EOD | 0.21^^^  | 0.17^^^  | 0.26^^^  |

Markit versus Fenics GG Price Discovery Metrics

|        | Total    | Europe   | US       |
|--------|----------|----------|----------|
| Markit | 0.72***  | 0.75***  | 0.69***  |
| Fenics | 0.28^^^  | 0.25^^^  | 0.31^^^  |

Markit versus Reuters EOD GG Price Discovery Metrics

|             | Total    | Europe   | US       |
|-------------|----------|----------|----------|
| Markit      | 0.78***  | 0.86***  | 0.70***  |
| Reuters EOD | 0.22^^^  | 0.14^^^  | 0.30^^^  |

Fenics versus Reuters EOD GG Price Discovery Metrics

|             | Total | Europe | US       |
|-------------|-------|--------|----------|
| Fenics      | 0.440 | 0.520  | 0.35^^^  |
| Reuters EOD | 0.560 | 0.480  | 0.65**   |

**Table 7: Inconsistency of the CDS databases based on a price discovery analysis between CDS spread returns and stock returns**

This table reports the results for the proximity and inconsistency of the different CDS databases based on a Granger causality test that is applied to pairs formed by the stock returns and CDS returns obtained from the quotes of four different CDS data sources (CMA, Markit, Reuters and Fenics). The sample comprises the Lehman Brothers collapse and spans from March 2008 to April 2009. Panel A reports the price discovery results summarized as the percentage of cases within each possible type of causality relation: (1) CDS returns cause stock returns and stock returns do not cause CDS returns, (2) CDS returns do not cause stock returns and stock returns cause CDS returns, (3) CDS returns cause stock returns and stock returns cause CDS returns, or (4) CDS returns do not cause stock returns and stock returns do not cause CDS returns. Panel B reports the average inconsistency across all 89 firms for all the possible pairs of data sources and for each individual pair of data sources. The average inconsistency measure for a given firm is defined as the sum of the cases in which the different pairs of databases are not in agreement in the causality relation between equity and CDS price discovery results relative to the total number pairs that can be formed with databases with information on the firm. The inconsistency measure for each pair of data sources take value one if the two datasets are in agreement in the causality relation and zero otherwise. \*\*\*, \*\*, and \* (^^^, ^^, and ^) denote whether the estimation is statistically different from 100% (0%) at 1, 5, and 10% significance levels.

| Panel A | CMA | Markit | Reuters | Fenics |
|---|---|---|---|---|
| Stocks cause CDS but CDS do not cause stocks | 19% | 18% | 23% | 39% |
| CDS cause stocks but stocks do not cause CDS | 1% | 3% | 5% | 3% |
| Double causation | 3% | 4% | 6% | 5% |
| No causation | 77% | 76% | 66% | 53% |
| Total | 100% | 100% | 100% | 100% |

| Panel B | | | |
|---|---|---|---|
| All pairs | 28% | *** | ^^^ |
| CMA - Markit | 11% | ** | ^^ |
| CMA - Reuters | 27% | *** | ^^^ |
| CMA - Fenics | 35% | *** | ^^^ |
| Markit - Reuters | 24% | *** | ^^^ |
| Markit - Fenics | 38% | *** | ^^^ |
| Fenics - Reuters | 33% | *** | ^^^ |

**Table 8: Determinants of the deviation
among the CDS data sources using alternative combinations of datasets**

This table reports the regression coefficients for the potential drivers of the dispersion among data sources using alternative combinations of these sources. The dependent variable is the logarithm of the standard deviation among the different CDS data sources employed in the analysis. The database includes 89 European and US firms (46 of the firms are European and 43 are American) which are the most liquid CDSs included in either the Itraxx or the CDX Index since the launching of the indexes, from January 2004 to March 2010. We estimate the coefficients for the above factors using a fixed-effects estimation procedure with standard errors clustered by firm and time (two-way cluster) and robust to heteroskedasticity. Column 1 reports the results for the whole sample of firms and is equivalent to the baseline regression estimation (Column 1 of Table 3); Column 2 reports the estimation results obtained when excluding Fenics to calculate the dependent variable and using information from CMA, Markit, and Reuters EOD; Column 3 reports the estimation results obtained when excluding Reuters EOD to calculate the dependent variable and using information from CMA, Markit, and Fenics. Each column contains the explanatory variables' coefficients and the standard errors between brackets. *** (** and *) indicates whether the coefficients are significant at a significance level of 1% (5% and 10%).

|   | (1) | (2) | (3) |
|---|---|---|---|
| Size (logarithm of market cap.) | -0.181** | -0.250** | -0.189** |
|  | (0.09) | (0.11) | (0.09) |
| Trade | -0.275*** | -0.146*** | -0.227*** |
|  | (0.04) | (0.04) | (0.04) |
| Days w/o a trade | 0.0008*** | 0.0004** | 0.0008*** |
|  | (0.00) | (0.00) | (0.00) |
| CDS Bid-Ask Spread | 0.054*** | 0.058*** | 0.053*** |
|  | (0.02) | (0.02) | (0.02) |
| Coefficient of variation of analysts' forecasts of EPS | 0.0002*** | 0.0002*** | 0.0002*** |
|  | (0.00) | (0.00) | (0.00) |
| Equity volatility (squared stock returns) | -0.645 | -0.787 | -0.425 |
|  | (0.86) | (1.01) | (0.76) |
| Financing costs of financial institutions | 0.496*** | 0.451*** | 0.415*** |
|  | (0.05) | (0.05) | (0.05) |
| Global risk (common implied volatility) | 0.025*** | 0.026*** | 0.023*** |
|  | (0.00) | (0.00) | (0.00) |
| Constant | 3.903* | 4.777* | 4.086** |
|  | (2.21) | (2.57) | 2.08 |
| R-squared | 0.430 | 0.290 | 0.440 |
| Number of observations | 132550 | 130219 | 132566 |
| Number of groups | 89 | 89 | 89 |
| F-statistic | 4228.370 | 3465.960 | 2980.380 |
| Prob. > F-statistic | 0 | 0 | 0 |
| Condition Index | 3.53 | 3.52 | 3.53 |

**Table 9: Determinants of the deviation among the CDS data sources using other estimation methodology and data frequency**

This table reports the regression coefficients for the potential drivers of the dispersion among data sources when using time fixed effects besides firm fixed effects and weekly instead of daily frequency. The dependent variable is the logarithm of the standard deviation among the different CDS data sources employed in the analysis. The database includes 89 European and US firms (46 of the firms are European and 43 are American) which are the most liquid CDSs included in either the Itraxx or the CDX Index since the launching of the indexes, from January 2004 to March 2010. Column 1 reports the baseline estimation results on the basis of a fixed-effects estimation procedure with standard errors clustered by firm and time (two-way cluster) and robust to heteroskedasticity. Column 2 reports the results obtained when we add time fixed effects to the estimation procedure employed in Colum 1. Instead of using a full set of time dummies we calculate the mean of the dependent variable on a daily basis and subtract this daily mean for each firm day by day. We exclude from this estimation global risk and financing costs variables. Column 3 reports the estimation results obtained when we use weekly instead of daily frequency. In this estimation we do not include the variable referred to the number of weeks without quotes and the dummy referred to the existence of trades through GFI in a given day is now transformed into a variable that takes values from 0 to 5 indicating the number of days with trades within a given week. Each column contains the explanatory variables' coefficients and the standard errors between brackets. *** (** and *) indicates whether the coefficients are significant at a significance level of 1% (5% and 10%).

|  | (1) | (2) | (3) |
|---|---|---|---|
| Size (logarithm of market cap.) | -0.181** | -0.113** | -0.160** |
|  | (0.09) | (0.05) | (0.08) |
| Trade | -0.275*** | -0.058** | -0.128*** |
|  | (0.04) | (0.03) | (0.02) |
| Days w/o a trade | 0.0008*** | 0.0006*** |  |
|  | (0.00) | (0.00) |  |
| CDS Bid-Ask Spread | 0.054*** | 0.021*** | 0.062** |
|  | (0.02) | (0.00) | (0.02) |
| Coefficient of variation of analysts' forecasts of EPS | 0.0002*** | 0.0001** | 0.0002*** |
|  | (0.00) | (0.00) | (0.00) |
| Equity volatility (squared stock returns) | -0.645 | 0.542 | -4.780 |
|  | (0.86) | (0.39) | (4.30) |
| Financing costs of financial institutions | 0.496*** |  | 0.488*** |
|  | (0.05) |  | (0.06) |
| Global risk (common implied volatility) | 0.025*** |  | 0.023*** |
|  | (0.00) |  | (0.00) |
| Constant | 3.903* | 2.559** | 3.296 |
|  | (2.21) | (1.17) | (2.22) |
| R-squared | 0.43 | 0.35 | 0.50 |
| Number of observations | 132550 | 132550 | 27556 |
| Number of groups | 89 | 89 | 89 |
| F-statistic | 4228.37 | 384.44 | 1142.52 |
| Prob. > F-statistic | 0 | 0 | 0 |
| Condition Index | 3.53 | 2.97 | 3.78 |

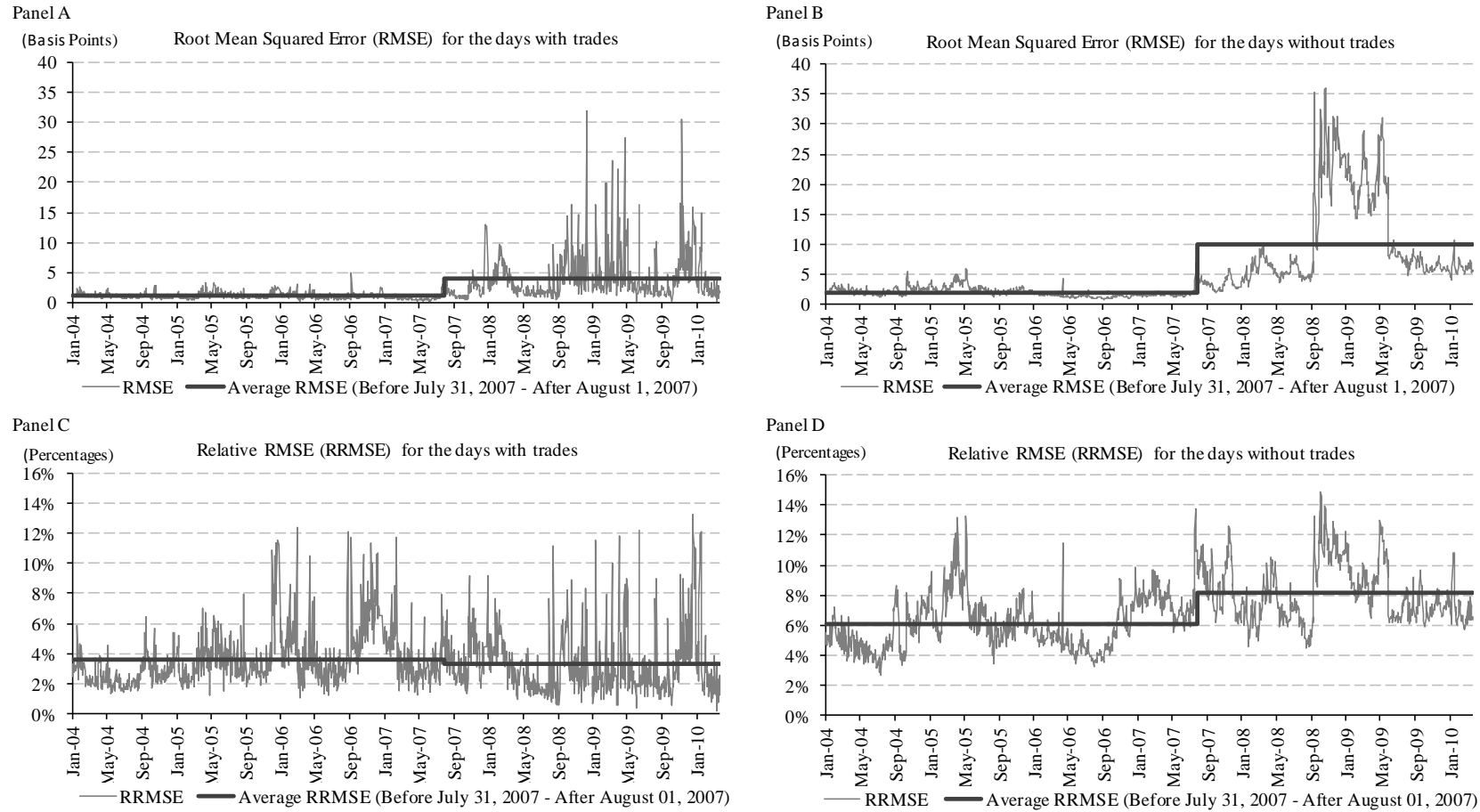

**Figure 1. Root Mean Squared Error (RMSE) and Relative Root Mean Squared Error (RRMSE).** This figure shows the cross-sectional deviations across data sources over time on the basis of two complementary indicators: Root Mean Squared Error (RMSE) in Panels A and B and Relative Root Mean Squared Error (RRMSE) in Panels C and D. The RMSE is obtained as the squared root of the average of the squared of the difference across pairs of data sources (CMA - Markit, CMA - Fenics, CMA - Reuters, Markit - Fenics, Markit - Reuters, Fenics – Reuters) for each firm every day. The RRMSE is obtained as the ratio of the RMSE to the average CDS spread across the four data sources for each firm every day. The time series in Figure 1 correspond to the daily cross-sectional average of the RMSEs and RRMSEs across the total number firms using the days in which there are trades for the firms (Panels A and C, respectively) and the days with no trades for the firms (Panels B and D, respectively), separately. We also report the average value of these time series for the pre-crisis (January 2004 – July 2007) and crisis (August 2007 – March 2010) periods.